\documentstyle[11pt,amssymbols]{article}
\title{A $q$-anaolg of the sixth Painlev\'e equation}
\author{
Michio Jimbo and Hidetaka Sakai\\
Department of Mathematics, Faculty of Science,\\
Kyoto University, Kyoto 606, Japan.
\cr}

\date{July 1995}

\begin{document}

\maketitle
\newtheorem{thm}{Theorem}
\newtheorem{prop}[thm]{Proposition}
\def\pf{\noindent{\it Proof.\quad}}
\def\qed{\hfil\fbox{}\medskip}
\def\displ#1{{\displaystyle #1}}
\def\Remark{\medskip\noindent {\sl Remark.}\quad}
\def\ga{\alpha}
\def\gep{\epsilon}
\def\gt{\theta}
\def\gk{\kappa}
\def\lgt{\Theta}
\def\cy{{\bf y}}
\def\cz{{\bf z}}
\def\cA{{\cal A}}
\def\diag{{\rm diag}}
\def\det{{\rm det}\,}

\def\hg(#1,#2;#3;#4){{}_2\phi_1\left({{{#1}\,\,\,{#2}}\atop{{#3}}};q,
                     {#4}\right)}

%%%%%%%%%%%%%%%%%%%%%%%%%%%%%%%%
\begin{abstract}
A $q$-difference analog of the
sixth Painlev\'e equation is presented.
It arises as the condition for preserving the connection matrix
of linear $q$-difference equations,
in close analogy with the monodromy preserving deformation
of linear differential equations.
The continuous limit and special solutions in terms of $q$-hypergeometric
functions are also discussed.
\end{abstract}
%%%%%%%%%%%%%%%%%%%%%%%%%%%%%%%%
\section{Introduction}

Recently
the intriguing idea of `singularity confinement' \cite{GRP}
has led to interesting
developments in discrete integrable systems.
It was introduced as a discrete counterpart of the Painlev\'e property.
As is well known,
the latter was the leading principle
in the classification of the Painlev\'e equations.
%have been isolated the of the by requiring this property.
In the same spirit, the singularity confinement test has
led to the discovery of difference analogs of several types of the
Painlev\'e equations \cite{RGH}.
%including a $q$-difference version of Painlev\'e III ($q$-$P_{III}$)
To our knowledge, difference versions of the Painlev\'e equations
are known except for the sixth type $P_{VI}$.

Another important aspect of the Painlev\'e equations is
their connection to monodromy preserving deformation of linear
differential equations.
Already in the classic paper of Birkhoff \cite{Bir},
the generalized Riemann problem
was studied for linear differential,
difference and $q$-difference equations in parallel.
An obvious next step would be to discuss the difference or $q$-difference
version of the deformation theory.
However we have been unable to find
such an attempt in the literature.
In the present article we report a simple non-trivial example
of this problem.
Namely we study the deformation of
a $2\times 2$ matrix system of linear $q$-difference
equations analogous to the linear differential equaitons underlying $P_{VI}$.
As a result we find a first order system of $q$-difference
equations, which we call $q$-$P_{VI}$ equation
(see (\ref{eqn:yy})--(\ref{eqn:zz})).
We shall also discuss some features of $q$-$P_{VI}$.

The text is organized as follows.
In Section 2 we recall known results concerning the analytic theory
of linear $q$-difference equations.
In Section 3 we illustrate their deformations
on the particular example mentioned above, and
derive a linear $q$-difference system with respect to the deformation
parameter.
The compatibility condition between the original and the
deformation equations
is worked out in Section 4, where
we find the $q$-$P_{VI}$ equation.
We show in Section 5 that it reduces
in the continuous limit $q\rightarrow 1$ to
a first order system equivalent to the original $P_{VI}$.
In Section 6 we discuss special solutions
given in terms $q$-hypergeometric functions,
which exist for special choice of parameters.
The final Section is devoted to discussions.

\section{Linear $q$-difference systems}

In this section we recall briefly
the classical theory of linear $q$-difference equations \cite{Bir}
which will be used later.
Throughout this article we fix a complex number $q$ such that $0<|q|<1$.

Consider an $m\times m$ matrix system with polynomial coefficients
\begin{equation}
Y(qx)=A(x)Y(x), \qquad
A(x)=A_0+A_1x+\cdots +A_N x^N.
\label{eqn:qdf}
\end{equation}
More general case of a rational $A(x)$ can be reduced to this case
by solving scalar $q$-difference equations.
We assume $A_0, A_N$ are semisimple and invertible.
Denoting by $\theta_j$, $\kappa_j$ ($1\le j\le m$) the eigenvalues of
$A_0$ and $A_N$ respectively, we assume further that
\[
\frac{\theta_j}{\theta_k}, \frac{\kappa_j}{\kappa_k}
{}~\not\in~\{q,q^2,q^3,\cdots\}
\qquad (\forall j, k).
\]
Set $A_0=C_0q^{D_0}C_0^{-1}$,
$A_\infty=C_\infty q^{D_\infty}C_\infty^{-1}$,
where
$D_0=\diag(\log\theta_j/\log q)$, $D_\infty=\diag(\log\kappa_j/\log q)$.

\begin{prop}[\cite{Bir}]
Under the conditions above, there exist unique solutions
$Y_0(x)$, $Y_\infty(x)$ of (\ref{eqn:qdf})
with the following properties:
\begin{eqnarray}
Y_0(x)&=&\widehat{Y}_0(x)x^{D_0},
\label{eqn:y0}\\
Y_\infty(x)&=&q^{\frac{N}{2}u(u-1)}\widehat{Y}_\infty(x)x^{D_\infty},
\qquad
u=\frac{\log x}{\log q}.
\label{eqn:yinf}
\end{eqnarray}
Here $\widehat{Y}_0(x)$ (resp. $\widehat{Y}_\infty(x)$) is
a holomorphic and invertible matrix at $x=0$ (resp. at $x=\infty$)
such that $\widehat{Y}(0)=C_0$ (resp. $\widehat{Y}_\infty(\infty)=C_\infty$).
\end{prop}

Let $\alpha_j$ ($j=1,\cdots,mN$) denote the zeroes of $\det A(x)$.
The $q$-difference equation (\ref{eqn:qdf}) entails
that $\widehat{Y}_\infty(x)^{\pm 1}$, $\widehat{Y}_0(x)^{\pm 1}$
can be continued meromorphically in the domain $0<|x|<\infty$.
Moreover
$\widehat{Y}_\infty(x)$ and $\widehat{Y}_0(x)^{-1}$ have no poles, while
$\widehat{Y}_\infty(x)^{-1}$ and $\widehat{Y}_0(x)$
are holomorphic except for possible poles at
\begin{eqnarray}
\widehat{Y}_\infty(x)^{-1} &:& q\alpha_j,q^2\alpha_j,q^3\alpha_j,\cdots,
\label{eqn:pole1}\\
\widehat{Y}_0(x) &:& \alpha_j,q^{-1}\alpha_j,q^{-2}\alpha_j,\cdots.
\label{eqn:pole2}
\end{eqnarray}

The connection matrix $P(x)$ is introduced by
\begin{equation}
Y_\infty(x)=Y_0(x)P(x).
\label{eqn:P}
\end{equation}
Clearly  $P(qx)=P(x)$.
It is known to be expressible in terms of elliptic theta functions.
It plays a role analogous to that of
the monodromy matrices for differential equations.

\section{Connection preserving deformation}

In the theory of monodromy preserving deformation of linear
differential equations,
one introduces extra parameter(s) $t$ in the coefficient matrix and
demand that the monodromy stay constant with respect to $t$.
Analogously, in the setting of $q$-difference equations,
one demands that the connection matrix stay
pseudo-constant in $t$, namely that $P(x,qt)=P(x,t)$.
The natural candidate for the deformation parameters
are the exponents
$\theta_j,\kappa_j$ at $x=0,\infty$ and the zeroes of $\det A(x)$.
(Notice that, unlike the case of
Fuchsian linear differential equations on ${\bf P}^1$,
the points $x=0,\infty$ play distinguished roles
in the $q$-difference systems.)
Under appropriate conditions it can be shown that
$P(x,t)$ is pseudo-constant in $t$ if and only if the corresponding solutions
$Y(x,t)=Y_0(x,t),Y_\infty(x,t)$ satisfy
\begin{equation}
Y(x,qt)=B(x,t)Y(x,t), \label{eqn:B}
\end{equation}
where $B(x,t)$ is
rational in $x$ (see Proposition \ref{prop:1} below).

{}From now on, we will
focus attention to the concrete example of a $2\times 2$ system
which is relevant to the $q$-$P_{VI}$ equation.
Recall that the linear system of differential equations associated with
$P_{VI}$ has the form \cite{JM}
\[
\frac{d}{dx}Y(x)=\cA(x)Y(x),
\qquad
\cA(x)=\frac{\cA_0}{x}+\frac{\cA_1}{x-1}+\frac{\cA_t}{x-t}.
\]
If one na\"{\i}vely replaces $d/dx$ by the $q$-differentiation symbol
$D_q=(1-q^{\vartheta})/(1-q)x$ ($\vartheta=xd/dx$) and multiplies through
the denominator, one obtains a $q$-difference system (\ref{eqn:qdf})
with
\begin{equation}
A(x)=(x-1)(x-t)\left(1-\epsilon x \cA(x)\right)
=A_0+A_1x+A_2x^2
\qquad (\epsilon=1-q).
\label{eqn:cA}
\end{equation}
Here
$A_2=I+\epsilon\cA_\infty$
($\cA_\infty=-\cA_0-\cA_1-\cA_t$) is independent of $t$, whereas
$A_0=t\left(I-\epsilon\cA_0\right)$ is proportional to $t$.
Since $\det A(0)$ is divisible by $t^2$, it is natural to
assume that two of the zeroes of $\det A(x)$ are divisible by $t$.

Motivated by this observation, we now take $A(x,t)$ to be of the form
\begin{eqnarray}
&&A(x,t)=A_0(t)+A_1(t)x+A_2x^2,
\label{eqn:A}\\
&&A_2=\pmatrix{\kappa_1&0\cr 0&\kappa_2\cr},
\qquad
A_0(t) \hbox{ has eigenvalues } t\theta_1, t\theta_2,
\label{eqn:A1}\\
&&\det A(x,t)=\kappa_1\kappa_2(x-ta_1)(x-ta_2)(x-a_3)(x-a_4).
\label{eqn:A2}
\end{eqnarray}
Here the parameters $\kappa_j,\theta_j,a_j$ are independent of $t$.
Clearly we have
\[
\kappa_1\kappa_2\prod_{j=1}^4a_j=\theta_1\theta_2.
\]
In what follows we will normalize $Y_\infty(x)$ by
$\widehat{Y}_\infty(\infty)=I$.

\begin{prop}\label{prop:1}
We have $P(x,qt)=P(x,t)$ if and only if (\ref{eqn:B})
holds for $Y=Y_0,Y_\infty$,
where $B(x,t)$ is a rational function of the form
\begin{equation}
B(x,t)=\frac{x\left(xI+B_0(t)\right)}{(x-qta_1)(x-qta_2)}.
\label{eqn:BB}
\end{equation}
\end{prop}
\begin{pf}
{}From the definition (\ref{eqn:P}), the connection matrix is
pseudo-constant in $t$ if and only if
\[
B(x,t)~{\buildrel {\rm def}\over =}~
Y_\infty(x,qt)Y_\infty(x,t)^{-1}=Y_0(x,qt)Y_0(x,t)^{-1}.
\]
Using (\ref{eqn:pole1}),(\ref{eqn:pole2}),
we find that the only poles in $0<|x|<\infty$
common to both sides are $x=qta_i$ ($i=1,2$).
Moreover (\ref{eqn:yinf}) along with the normalization of $Y_\infty(x)$
imply that the left hand side behaves as $I+O(x^{-1})$ at $x=\infty$.
Similarly (\ref{eqn:y0}) implies that the right hand side behaves like $O(x)$
at $x=0$ (notice that $D_0$ is proportional to $t$).
The proposition is an immediate consequence of these properties.
\qed
\end{pf}

\section{Derivation of $q$-$P_{VI}$}
The compatibility condition for the systems
(\ref{eqn:qdf}), (\ref{eqn:B}) reads
\begin{equation}
A(x,qt)B(x,t)=B(qx,t)A(x,t)
\label{eqn:comp}
\end{equation}
where $A(x,t)$ and $B(x,t)$ are given respectively by
(\ref{eqn:A}) and (\ref{eqn:BB}).
We will now work out the implications of (\ref{eqn:comp}) and
find the $q$-$P_{VI}$ equation.

Define $y=y(t)$, $z_i=z_i(t)$ ($i=1,2$) by
\begin{equation}
A_{12}(y,t)=0, \qquad A_{11}(y,t)=\kappa_1z_1,
\qquad A_{22}(y,t)=\kappa_2z_2,
\label{eqn:yzz}
\end{equation}
so that $z_1z_2=(y-ta_1)(y-ta_2)(y-a_3)(y-a_4)$.
In terms of $y,z_1,z_2$ and (\ref{eqn:A})--(\ref{eqn:A2}),
the matrix $A(x,t)$ can be parametrized as follows.
\begin{eqnarray*}
&&A(x,t)=\left(\matrix{
\kappa_1\bigl((x-y)(x-\alpha)+z_1\bigr)&\kappa_2w(x-y)\cr
\kappa_1w^{-1}(\gamma x+\delta)&\kappa_2\bigl((x-y)(x-\beta)+z_2\bigr)
\cr}\right).
\end{eqnarray*}
Here
\begin{eqnarray*}
&&\alpha
={1\over\kappa_1-\kappa_2}[y^{-1}((\theta_1+\theta_2)t-\kappa_1z_1-\kappa_2z_2)
-\kappa_2((a_1+a_2)t+a_3+a_4-2y)], \\
&&\beta={1\over\kappa_1-\kappa_2}[-y^{-1}((\theta_1+\theta_2)t-\kappa_1z_1-\kappa_2z_2)
+\kappa_1((a_1+a_2)t+a_3+a_4-2y)],  \\
&&\gamma=z_1+z_2+(y+\alpha)(y+\beta)+(\alpha+\beta)y
-a_1a_2t^2-(a_1+a_2)(a_3+a_4)t-a_3a_4, \\
&&\delta=y^{-1}(a_1a_2a_3a_4t^2-(\alpha y+z_1)(\beta y+z_2)).
\end{eqnarray*}
The quantity $w=w(t)$ is related to the `gauge' freedom, and
does not enter the final result for the $q$-$P_{VI}$ equation.

The compatibility (\ref{eqn:comp}) is equivalent to
\begin{eqnarray}
&&A(qa_it,qt)\left(qa_itI+B_0(t)\right)=0\qquad(i=1,2)
\label{eqn:co1}\\
&&\left(qa_itI+B_0(t)\right)A(a_it,t)=0\qquad(i=1,2)
\label{eqn:co2}\\
&&A_0(qt)B_0(t)=qB_0(t)A_0(t).
\label{eqn:co3}
\end{eqnarray}
Substituting the parametrization above one obtains
a set of $q$-difference equations among the quantities $y,z_1$,etc.
We will not go into the details of the cumbersome calculation, but
merely state the result.

Following \cite{GNPRS} let us use the notations
\[
\overline{y}=y(qt),\qquad \underline{y}=y(q^{-1}t)
\]
and so forth.
Introduce $z$ by
\[
z_1=\frac{(y-ta_1)(y-ta_2)}{q\kappa_1 z},
\qquad
z_2=q\kappa_1 (y-a_3)(y-a_4)z.
\]
Then the matrix $B_0(t)=\left(B_{ij}\right)$ is parametrized as follows:
\begin{eqnarray*}
B_{11}&=&\frac{-q\kappa_2\overline{z}}{1-\kappa_2\overline{z}}
\left(-\beta+\frac{t(a_1+a_2)-y}{\kappa_2\overline{z}}\right),
\\
B_{22}&=&\frac{-q\kappa_1\overline{z}}{1-q\kappa_1\overline{z}}
\left(-\overline{\alpha}+\frac{tq(a_1+a_2)-\overline{y}}
{q\kappa_1\overline{z}}\right),
\\
B_{12}&=&\frac{q\kappa_2\overline{z}}{1-\kappa_2\overline{z}}w,
\\
B_{21}&=&\frac{q\kappa_1\overline{z}}{w(1-q\kappa_1\overline{z})}
\left(tqa_1-\overline{\alpha}+\frac{tqa_2-\overline{y}}
{q\kappa_1\overline{z}}\right)
\left(ta_1-\beta+\frac{ta_2-y}{\kappa_2\overline{z}}\right)
\\
&=&\frac{q\kappa_1\overline{z}}{w(1-q\kappa_1\overline{z})}
\left(tqa_2-\overline{\alpha}+\frac{tqa_1-\overline{y}}
{q\kappa_1\overline{z}}\right)
\left(ta_2-\beta+\frac{ta_1-y}{\kappa_2\overline{z}}\right).
\end{eqnarray*}

Set further
\begin{equation}
b_1=\frac{a_1a_2}{\theta_1},\quad
b_2=\frac{a_1a_2}{\theta_2},\quad
b_3=\frac{1}{q\kappa_1},\quad
b_4=\frac{1}{\kappa_2}.
\label{eqn:bj}
\end{equation}

\begin{thm}
The equations (\ref{eqn:co1})--(\ref{eqn:co3}) are equivalent to
\begin{eqnarray}
\frac{y\overline{y}}{a_3a_4}&=&
\frac{(\overline{z}-tb_1)(\overline{z}-tb_2)}
{(\overline{z}-b_3)(\overline{z}-b_4)},
\label{eqn:yy}\\
\frac{z\overline{z}}{b_3b_4}&=&
\frac{(y-ta_1)(y-ta_2)}{(y-a_3)(y-a_4)},
\label{eqn:zz}\\
\frac{\overline{w}}{w}&=&
\frac{b_4}{b_3}.
\frac{\overline{z}-b_3}{\overline{z}-b_4}.
\label{eqn:ww}
\end{eqnarray}
Here $b_j$'s are given by (\ref{eqn:bj}). We have a single constraint
\[
\frac{b_1b_2}{b_3b_4}=q\frac{a_1a_2}{a_3a_4}.
\]
\end{thm}

We call (\ref{eqn:yy})--(\ref{eqn:zz}) $q$-$P_{VI}$ system, or simply
$q$-$P_{VI}$ equation.
Note that  the number of parameters can be reduced to $4$
by rescaling $y,z,t$;
e.g. one can choose $a_1a_2=1,a_3a_4=1,b_1b_2=q,b_3b_4=1$.
Written in the first order form,
the map $(y,z)\mapsto (\overline{y},\overline{z})$ is birational.
Upon elimination of $z$, however, $\overline{y}$ becomes
double-valued as a function of $y$ and $\underline{y}$.

One can verify without difficulty
that the $q$-$P_{VI}$ system (\ref{eqn:yy})--(\ref{eqn:zz})
possesses the singularity confinement property in  the sense of
\cite{GRP,RGH}.
At this moment we do not know how it is related to
the other discrete $P_{V}$--$P_I$ equations
(see \cite{RGH} and references therein).

\Remark
It is worth mentioning that the $q$-$P_{III}$ equation \cite{RGH}
has a very similar form
\[
\frac{\underline{w}\overline{w}}{a_3a_4}
=
\frac{(w-ta_1)(w-ta_2)}{(w-a_3)(w-a_4)}
\]
where $a_1,\cdots,a_4$ are arbitrary parameters.
In fact the linear problem
given in \cite{GNPRS,KOS} falls within the present framework.
To see this, set $Y=D\Phi$ with $D=\diag(1,1,h^2,h^2)$
in the notation of \cite{KOS},
and rename the parameters $q^{2}$, $x^{2}$ and $h^{2}$ there
by $q$, $t$ and $x$.
The linear system for $q$-$P_{III}$ then takes the form (\ref{eqn:qdf}),
(\ref{eqn:B}) with
\[
A(x,t)=\frac{1}{x}A_0(t)+A_1(t),\qquad B(x,t)=\frac{1}{x}B_0(t)+B_1(t),
\]
where $A_j,B_j$ are $4\times 4$ matrices given as follows.
\begin{eqnarray*}
&&
A_0=\pmatrix{ & &\alpha                      & 0         \cr
              & &q^{-1}\underline{\beta}+\tau/t&q^{-1}\underline{\beta}\cr
\beta  & 0   & & \cr
q^{-1}\underline{\alpha}+\tau/t&q^{-1}\underline{\alpha} &  & \cr},
\qquad
A_1=\pmatrix{\kappa&\kappa+\alpha&    &          \cr
             0     &\tau/t         &    &          \cr
                   &             &\xi &\xi+\beta \cr
                   &             & 0  &\tau/t      \cr},
\\
&&
B_0=\pmatrix{ & & 0   & 0         \cr
              & & 1   & 0         \cr
             0&0&     &           \cr
             1&0&     &           \cr},
\qquad
B_1=\pmatrix{
\displ{\frac{\kappa-\tau/t}{\alpha+\tau/t}}
&\displ{\frac{\kappa+\alpha}{\tau/t+\alpha}}&  &  \cr
             0     & 0         &    &          \cr
  &  &
\displ{\frac{\xi-\tau/t}{\beta+\tau/t}}
&\displ{\frac{\xi+\beta}{\tau/t+\beta}}\cr
                   &             & 0  &0      \cr}.
\end{eqnarray*}
Here
$\kappa=-qa_1a_4/a_3$, $\xi=-qa_2$, $\tau=-qa_4$,
$\alpha=qa_1a_4/w$, $\beta=qw/t$.
The eigenvalues of $A_i$ are
\begin{eqnarray*}
A_0 &:& \pm c t^{-1/2}, \pm cq^{1/2}t^{-1/2} \qquad(c=q\sqrt{a_1a_4}) \\
A_1&:& \kappa,\xi,\tau t^{-1},\tau t^{-1},
\end{eqnarray*}
and $\det A(x,t)$ has zeroes at $x^2=1, -qa_2/a_1a_3$.
Note that in this example the exponents at $x=0,\infty$ are moving
with resptect to $t$ while the zeroes of  $\det A(x,t)$ are fixed.
\qed

\section{Continuous limit}

{}From the construction one expects that in the continuous limit
the $q$-$P_{VI}$ equation reduces to the $P_{VI}$ differential equation.
Here by continuous limit we mean the limit $\gep=1-q \rightarrow 0$.
In view of the relation (\ref{eqn:cA}) it is natural to set
\[
\gk_i=1+\gep K_i, \quad
\gt_i=1-\gep\lgt_i,  \quad
a_i=1+\gep\ga_i.
\]
Note that
\[
(1+\gep K_1)(1+\gep K_2)
(1+\gep \ga_1)(1+\gep \ga_2)(1+\gep \ga_3)(1+\gep \ga_4)
=(1-\gep \lgt_1)(1-\gep \lgt_2).
\]
Redefinig $\cy=y$ and $\cz$ by
\[
z_1={1\over\gk_1}(y-t)(y-1)(1-\gep y {\cz})
 \]
we find
\begin{eqnarray}
\frac{d{\cy}}{dt}&=&
{\cy(\cy-1)(\cy-t)\over t(t-1)}
\left(2\cz-\frac{\lgt_1+\lgt_2}{\cy}-
\frac{\ga_3+\ga_4}{\cy-1}
-\frac{\ga_1+\ga_2-1}{\cy-t}\right)
\label{PVI1} \\
\frac{d\cz}{dt}&=&
\frac{-3\cy^2+2(t+1)\cy-t}{t(t-1)}\cz^2\nonumber \\
&&+\frac{(2\cy-t-1)(\lgt_1+\lgt_2)+(2\cy-1)(\ga_1+\ga_2-1)+(2\cy-t)(\ga_3+\ga_4)}
{t(t-1)}\cz\nonumber \\
&&-\frac{K_1(K_2+1)}{t(t-1)}
+\frac{\lgt_1\lgt_2}{(t-1)\cy^2}+\frac{\ga_1\ga_2}{(\cy-t)^2}
-\frac{\ga_3\ga_4}{t(\cy-1)^2}
\label{PVI2}.
\end{eqnarray}
This is a first order system equivalent to the
$P_{VI}$ differential equation.

\section{Special solutions}

At particular values of parameters,
the system (\ref{eqn:yy})--(\ref{eqn:zz}) decouples into a pair of
$q$-Riccati equations, in exactly the same way as for
the other discrete Painlev\'e equations \cite{GNPRS,PNGR,KOS}.
Namely, assume that
\[
\frac{b_1}{b_3}=q\frac{a_1}{a_3},
\qquad
\frac{b_2}{b_4}=\frac{a_2}{a_4}.
\]
Then (\ref{eqn:yy}), (\ref{eqn:zz}) are satisfied if
\[
\overline{y}=a_3\frac{\overline{z}-tb_1}{\overline{z}-b_3},
\qquad
\overline{z}=b_4\frac{y-ta_2}{y-a_4}.
\]
The latter can be linearized in the standard way.
Let
\[
t=\frac{b_3}{b_1}s,
\quad
a=\frac{a_3}{a_4},
\quad
b=\frac{a_2}{a_4}\frac{b_4}{b_1},
\quad
c=\frac{a_3}{a_4}\frac{b_4}{b_3}.
\]
Then
\[
y=a_3\frac{u}{v},\qquad
\overline{z}=b_4\frac{u-(bs/c)v}{u-v/a}
\]
where $u=u(s),v=v(s)$ are solutions of
\begin{eqnarray}
\overline{u}&=&
\frac{1}{1-(ab/c)s}\left((1-\frac{a}{c}s)u+\frac{1-b}{c}sv\right),
\label{eqn:hyp1}\\
\overline{v}&=&
\frac{1}{1-(ab/c)s}\left((1-\frac{a}{c})u+\frac{(1-bs)}{c}v\right).
\label{eqn:hyp2}
\end{eqnarray}
In particular the $q$-hypergeometric functions
\[
u=\hg(a,b;c;s),
\qquad
v=\frac{c-a}{c-1}\,\hg(a,qb;qc;s)
\]
solve (\ref{eqn:hyp1})--(\ref{eqn:hyp2}).

\section{Discussions}

In this note we studied a deformation of a linear $q$-difference system,
which led to the $q$-$P_{VI}$ equation.
The argument presented here has a very general character.
We feel the subject warrants further investigation to develop
a general theory of deformation in the difference/$q$-difference setting,
including $\tau$-functions,
symplectic structure,
Schlesinger transforms and symmetries.
Another interesting problem is to explore an
analog of Okamoto's space of initial conditions \cite{O1},
in connection with the affine Weyl group
symmetry of the Painlev\'e equaitons \cite{O2}.
This might shed light to the geometric meaning of the singularity confinement.

\vskip 1cm
\noindent{{\it Acknowledgements.}\quad}
We wish to thank
K. Aomoto,
B. Grammaticos,
K. Kajiwara,
J. Matsuzawa,
A. Ramani,
and
N. Yu.  Reshetikhin
for discussions and interest.
This work is partly supported by Grant-in-Aid for Scientific Research
on Priority Areas 231, the Ministry of Education, Science and Culture.

\bibliographystyle{unsrt}

\end{document}